\documentclass{article}

\usepackage[a4paper, margin=0.8in]{geometry}
\usepackage[utf8]{inputenc}
\usepackage[dvipsnames,table,xcdraw]{xcolor}
\usepackage[affil-it]{authblk}
\usepackage[misc]{ifsym}
\usepackage{multirow}
\usepackage{listings}
\usepackage{float}
\usepackage{algorithm}
\usepackage{algorithmicx}
\usepackage{algpseudocode} 
\usepackage{adjustbox}
\usepackage{tikz}
\usetikzlibrary{quantikz2}

\usepackage{appendix,verbatim,listings,braket,enumitem, kantlipsum,graphicx,multicol,mathtools,csquotes,amsmath,mathtools,url,subfiles,amsthm,amssymb,fdsymbol,rotating,caption,hyperref,fontawesome,xcolor,arydshln,ulem,subcaption,caption, amsfonts}
\usepackage{longtable}
\usepackage{pgfplots}
\usepackage{tikz}
\usetikzlibrary{shapes.geometric, arrows, positioning, backgrounds, shapes.arrows}
\usepackage{diagbox}
\usepackage{setspace}

\usepackage{xspace}
\newcounter{mynote}

\usepackage{sectsty}
\def\titlefont{\color{RoyalPurple}}
\sectionfont{\color{RoyalPurple}}
\subsectionfont{\color{RoyalPurple}}
\subsubsectionfont{\color{RoyalPurple}}

\let\OLDthebibliography\thebibliography
\renewcommand\thebibliography[1]{
  \small
  \OLDthebibliography{#1}
 \setlength{\itemsep}{0pt minus 0.3ex}
}

\title{\titlefont \textbf{\Large DeQompile: quantum circuit decompilation using\\genetic programming for explainable quantum architecture search}}

\author[1,2]{Shubing Xie}
\author[2,3]{Aritra Sarkar}
\author[2,3 \Letter]{Sebastian Feld}

\affil[1]{Instituut-Lorentz, Leiden University, The Netherlands}
\affil[2]{Quantum Machine Learning research group,
Quantum Computing division, QuTech, The Netherlands}
\affil[3]{Department of Quantum \& Computer Engineering, Delft University of Technology, The Netherlands}

\affil[ \Letter ]{s.feld@tudelft.nl}

\date{}

\begin{document}

\maketitle

\begin{abstract}
Demonstrating quantum advantage using conventional quantum algorithms remains challenging on current noisy gate-based quantum computers. 
Automated quantum circuit synthesis via quantum machine learning has emerged as a promising solution, employing trainable parametric quantum circuits to alleviate this.
The circuit ansatz in these solutions is often designed through reinforcement learning-based quantum architecture search when the domain knowledge of the problem and hardware are not effective. 
However, the interpretability of these synthesized circuits remains a significant bottleneck, limiting their scalability and applicability across diverse problem domains.

This work addresses the challenge of explainability in quantum architecture search (QAS) by introducing a novel genetic programming-based decompiler framework for reverse-engineering high-level quantum algorithms from low-level circuit representations. 
The proposed approach, implemented in the open-source tool DeQompile, employs program synthesis techniques, including symbolic regression and abstract syntax tree manipulation, to distill interpretable Qiskit algorithms from quantum assembly language. 
Validation of benchmark algorithms demonstrates the efficacy of our tool. 
By integrating the decompiler with online learning frameworks, this research potentiates explainable QAS by fostering the development of generalizable and provable quantum algorithms.
\end{abstract}

\section{Introduction} \label{section:intro}


Quantum computing (QC) represents a paradigm shift in information processing, utilizing the principles of quantum mechanics to solve problems that are intractable for classical computers. 
In recent years, significant advances have been made in engineering gate-based quantum computing, which relies on quantum bits (qubits) and quantum gates to manipulate quantum states and process data in a superposition of states \cite{nielsen2010quantum}. 
The current state of the art in quantum computing is exemplified by breakthroughs from quantum processor manufacturers in qubits with better fidelity, scalability, real-time control, and error-correction \cite{acharya2024quantum, kim2023evidence, akahoshi2024partially}.
Conventional quantum algorithms like integer factorization \cite{shor1999polynomial} and search \cite{grover1996fast} ensure provable resource complexity advantages over their classical counterparts.
However, demonstrating the classical-to-quantum advantage crossover for applications via these algorithms requires considerable improvements in the fidelity and multiplicity of qubits. 
The current noisy intermediate-scale quantum computing (NISQ) \cite{preskill2018quantum, leymann2020bitter, waintal2024quantum, ezratty2023we} devices have motivated the automated synthesis of quantum circuits while accounting for the device constraints \cite{chong2023closing, zimboras2025myths}.

The quantum circuit synthesis methods are often referred to using the broader umbrella term of quantum machine learning (QML).
Similar to classical neural networks, these methods typically consider a parametric quantum circuit (PQC)  \cite{mcclean2016theory, cerezo2021variational}.
The structure/ansatz of the circuit can either be based on the problem structure or the hardware constraints or automatically constructed via quantum architecture search (QAS) \cite{du2022quantum, zhu2022brief, martyniuk2024quantum}.
The parameters, in turn, are iteratively adjusted by the variational principle mediated by a classical optimizer in the loop.
QAS is typically mediated by neural-network-based reinforcement learning (RL-QAS) \cite{ostaszewski2021reinforcement, kundu2024reinforcement} or population-based evolutionary algorithms \cite{ding2022evolutionary, zhang2023evolutionary}.

QAS-based PQC shares both the strengths and limitations of classical machine learning (ML) based neural architecture search (NAS).
A major challenge in deep neural networks has been the issue of explainability \cite{gilpin2018explaining} of both the architecture and the learned model.
Various model explainability techniques have been developed in ML and subsequently in QML \cite{heese2023explaining, gil2024opportunities, pira2024interpretability, power2024feature, lin2024quantum}.
Interpretable NAS techniques \cite{ru2020interpretable, pereira2023neural} have also been employed for post-doc understanding of the architecture based on empirical performance.
Similarly to NAS, while QAS-based solutions have been shown to perform against hardware constraints \cite{patel2024curriculum}, the trained ansatz is not human-interpretable.
This restricts the generalizability of these solutions to larger problem sizes and a deeper understanding of the underlying governing symmetries that led the QAS to converge on the solution.
Recently, explainable QAS has been attempted variously via quantum information theoretic approach \cite{sadhu2024quantum}, interpretable learning models \cite{kundu2024kanqas}, and online learning of a gadget library (GRL-QAS) \cite{kundu2024easy}.
In this work, we develop a complementary approach toward explaining a family of quantum circuits via a genetic programming-based decompiler.
We demonstrate our framework using examples from conventional quantum algorithms and suggest their integration within a GRL-QAS framework to augment their capability beyond simple gadgets.

Program synthesis provides a promising avenue for addressing the explainability problem in quantum algorithm design automation \cite{sarkar2024automated}. 
It aims to distill a family of circuits into a high-level, interpretable algorithm, often by using methods like genetic programming or neural networks. 
Symbolic regression, for example, can be used to find mathematical expressions that describe the behavior of quantum circuits \cite{schmidt2009distilling}.
For more holistic and expressive descriptions, high-level program synthesis and concept learning \cite{bowers2023top, trenkwalder2022automated} can be employed.
Decompilation closely resembles inductive inference in the human brain, such as when we deduce the next item in a sequence by identifying underlying patterns. 
In the context of quantum computing, such methods would facilitate the understanding and generalization of quantum algorithms across different quantum hardware and related problem instances.
For quantum circuits, if the circuits are known at a limited scale and we wish to extend the corresponding design, we must identify the shared structures and recognize the patterns of these circuits. 
Inspired by quantum techniques for Solomonoff's induction \cite{sarkar2022applications,sarkar2022qksa,sarkar2020quantum,sarkar2021estimating}, in this research, the rules or patterns of these circuits are represented by finding the underlying code that can generate them.
With that background, in this article, we address this research question of reverse-engineering high-level quantum algorithms (in Qiskit) from their low-level representations (QASM) using genetic programming (GP) on the abstract syntax tree (AST) representation.
We develop a framework to tackle this question, including a method to initialize syntactically valid Qiskit ASTs, a multi-objective fitness function, the genetic operators, and strategies to improve the convergence of the GP.
We validate the decompiler's performance on common variational ansatz and quantum algorithms like GHZ state preparation, quantum Fourier Transform, and quantum phase estimation.
Further, we investigate the increased difficulty in decompiling circuits that have been transpiled to the constraints of underlying quantum hardware.
The software implementation of our technique, called DeQompile, is available as open-source software for further research and development.


\section{Background} \label{sec:background}



This section introduces the necessary concepts of decompilation, genetic programming, and abstract syntax tree representation.

\subsection{Decompilation}

Decompilation is the process of translating compiled code (e.g., low-level assembly code) back into a more readable form of source code (e.g., a high-level language) or a close approximation of it without having access to the original source code.
This process is crucial for understanding, analyzing, and optimizing software systems, especially when the original source code is obfuscated or unavailable. 
In classical computing, decompilation has numerous applications, including software reverse engineering, explainable AI (XAI), and programming by example. 
For instance, in software reverse engineering, decompiling a program’s binary allows for vulnerability assessment, malware analysis, and system optimization. 
Decompilation plays an important role in programming by example, such as Excel's Flash Fill, where it is used to automate string processing tasks based on input-output examples, significantly reducing the need for manual coding \cite{flashfill, egele2008survey, royal2006polyunpack}.
In the realm of XAI, decompiling machine learning models \cite{hu2024degpt, blazek2024automated} help enhance interpretability and transparency, which is vital for ensuring the trustworthiness of AI systems. 


Decompilation, however, is inherently a difficult task due to several challenges.
Fundamentally, the undecidability of the halting problem implies that it is not always possible to determine whether a synthesized program will terminate with the intended effect or run indefinitely. 
Additionally, the program space is non-differentiable, which prevents the direct application of traditional gradient-based optimization methods. 
The semantic variable names are typically absent in low-level codes, making it difficult to understand the decompiled code.
Despite these difficulties, various techniques have been developed to address the challenges of decompilation. 
For instance, genetic programming \cite{Koza1992}, a technique rooted in evolutionary algorithms, has been popularly used for program synthesis and software reverse engineering. 
Neural networks \cite{baldoni2018survey} have also been applied to decompilation tasks, where deep learning models can learn mappings between machine code and high-level languages, thus automating and enhancing the decompilation process.

Despite the similarities with classical decompilation, quantum circuit decompilation has not yet been explored for quantum algorithm reverse engineering from learned quantum circuits.
Quantum decompilation faces unique challenges that differentiate it from classical decompilation. 
Quantum phenomena like superposition, entanglement, non-stabilizer magic, and interference do not have phenomenological analogs in human daily experience (or in propositional logic), thus requiring a certain degree of commitment to mathematical models in their interpretability.
It is also preferable to evaluate the decompiler purely syntactically instead of testing by executing the decompiler artifact to circumvent the exponential computational cost of classical simulation of quantum programs.
Besides this, quantum transformations have an added continuous yet non-measurable degree of freedom in their global phase, which makes it difficult to generalize across circuits where the pattern gets obscured by this phase factor.

Decompiling quantum circuits could uncover the underlying principles of quantum architecture and algorithm design, ultimately enabling the discovery of more interpretable, scalable, and provable quantum algorithms.
This would allow us to transition from QAS-based PQC solutions in the NISQ era to conventional quantum algorithms in the fault-tolerant quantum computing (FTQC) era.



\subsection{Genetic programming}

The basic idea behind evolutionary techniques such as genetic algorithms (GA), genetic programming (GP), and genetic expression programming (GEP) is to evolve solutions to problems by iteratively modifying a population of candidate solutions. 
These candidates are typically represented as tree structures, which can be subjected to operations like selection, crossover, and mutation to explore the solution space.
GP \cite{Koza1992} allows computer programs to evolve to solve a variety of complex problems, including symbolic regression, classification, and optimization. 
In GP, the nodes of the tree represent functions or operations, and the leaves represent variables or constants. 
The fitness function evaluates the performance of the individuals (programs) and guides the evolution process.
Quantum genetic programming (QGP) \cite{spector2004automatic, Spector2003} extends the concept of classical GP into the realm of quantum program synthesis.

The steps of initialization, evaluation, selection, crossover, and mutation characterize the GP methodology.
The following algorithm outlines the baseline genetic programming process.

\begin{algorithm}[H]
\caption{Baseline Genetic Programming Algorithm}
\begin{algorithmic}[1]
\State Initialize population $P$ with random individuals (programs in chosen representation)
\For{each generation $g$}
\State Evaluate the fitness of each program in $P$ based on a fitness function
\State Select the fittest individuals from $P$ to form a mating pool
\State Apply crossover to pairs of individuals in the mating pool to create offspring
\State Apply mutation to the offspring with a certain probability
\State Replace the least fit individuals in $P$ with the new offspring
\EndFor
\State Return the best individual from the final population
\end{algorithmic}
\end{algorithm}

A critical challenge in both classical and quantum GP is the encoding of the candidate solutions to allow efficient manipulation and evolution during the search process. 
This is where abstract syntax trees (AST) come into play and will be discussed in more detail subsequently.

\subsection{Abstract syntax tree}

Abstract syntax trees are a hierarchical, tree-like representation of the syntactic structure of source code or expressions, abstracted from the details of the underlying syntax. 
Unlike the raw code or textual representation, the AST focuses on the logical structure and relationships within the code, omitting unnecessary syntactic details like punctuation and formatting. 
Each node in the tree corresponds to a construct or operation in the source code, such as a variable, function, operator, or control flow structure.

ASTs are widely used in compilers, interpreters, and program analysis tools for efficient parsing, optimization, and transformation of code. 
AST's hierarchical and structured representation of quantum circuits or programs makes them ideal for encoding in evolutionary algorithms. 
By using ASTs, we can effectively manipulate the structure of quantum programs and optimize them during the evolutionary process, facilitating the exploration of the quantum program space.

\section{Quantum circuit decompilation}

In this section, we explain the decompilation process for QASM circuits.
A population of Qiskit programs is initialized in the AST representation.
The fitness for each corresponding Qiskit code for each AST is evaluated by comparing the generated QASM from the code and the training corpus (i.e., the list of QASM to be decompiled).
The GP guides the evolution towards fitter individuals that can decompile the corpus.

\subsection{Problem formulation}

We study two primary objectives.
Firstly, we select a quantum algorithm \( A \) of choice, such as simple ansatzes, GHZ state preparation, quantum Fourier transform, quantum phase estimation, etc. 
Quantum circuits for \( A \) are generated for different problem sizes, ranging from $2$ to $30$ qubits, using a Python program \( P_A(n) \) using the Qiskit package \cite{javadi2024quantum}. 
The resulting circuits, \( C_A^2, C_A^3, \dots, C_A^{30} \), is represented in OpenQASM \cite{cross2022openqasm}. 
This set of circuits \( C_A^n \) serves as the training data set. 
Given this set as input, the designed decompiler evolves an optimal program \( P'_A(n) \) that closely approximates \( P_A(n) \). 
The closeness metrics can be either the process distance between the synthesized unitaries or the difference between the generated QASMs, as discussed later.
\( P'_A(n) \) can then be used to generate \( {C'}_A^{31}, {C'}_A^{32}, \dots \).
The data set can also be divided into training and test sets to validate the generalization capability.
    
Secondly, we explore the limits of the tool through empirical analysis of decompiling a highly optimized code with a lesser algorithmic structure. 
We take circuits \( C_A^n \) and translate them into the corresponding unitaries \( U_A^n \). 
These unitaries are decomposed using Qiskit's transpilation into the native gate set of a quantum processing unit (QPU), such as IBM's, to obtain circuits \( {C'}_A^n \). 
Given this set of circuits \( {C'}_A^n \), the decompiler will infer \( P''_A(n) \).
We show our experiments agree with the expectation that \( P''_A(n) \) is less explainable and more abstract than \( P'_A(n) \), thus making it harder to generalize from optimized circuits. 

\begin{figure}[H]
    \centering 
    \includegraphics[clip, trim=3.2cm 0.0cm 0.0cm 2.3cm, width=\textwidth]{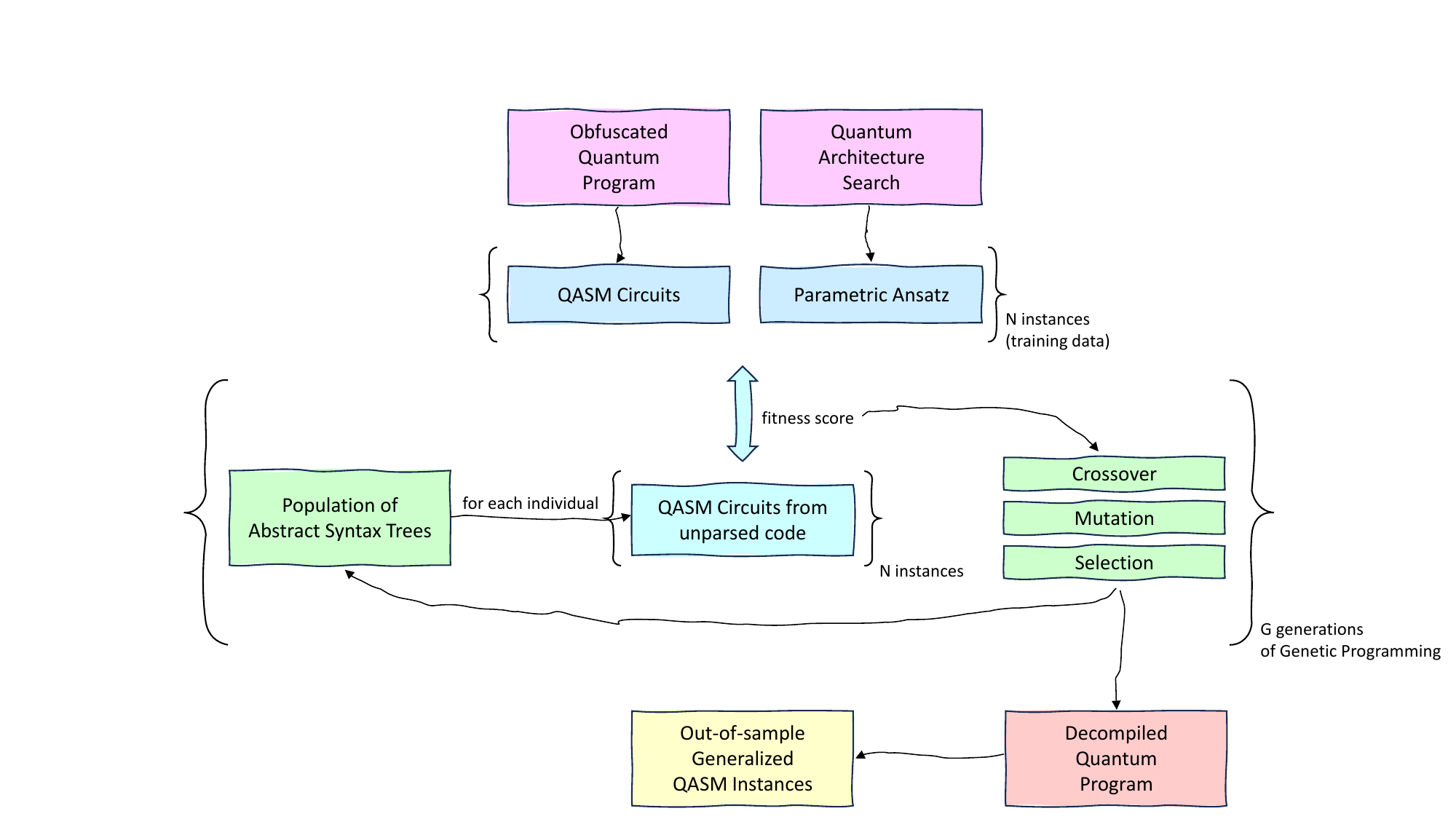}
    \caption{Software architecture of DeQompile}
    \label{Decompile}
\end{figure}

The software architecture of DeQompile is shown in the Figure \ref{Decompile}.


\subsection{AST representation}

In this project, we use the \texttt{ast} module of Python to generate a population of individuals for the GP.
The \texttt{parse} and \texttt{unparse} functions in the \texttt{ast} package allow the conversion between the AST and code string representation.
Each AST in the population represents a Python function that takes as an argument a circuit size $n$ and returns a Qiskit quantum circuit object of $n$ qubits.
The operations in the quantum circuit are (potentially) conditioned on $n$ as a proxy for the problem instance size.
Each node in the AST represents a fundamental construct in the source code, such as quantum gate operations, assignments, and control flow. 

The nodes of the AST include:
\begin{itemize}[nolistsep,noitemsep]
    \item Module: Represents the entire quantum program, containing all quantum operations and gates.
    \item FunctionDef: Represents a function definition, such as a quantum circuit initializer.
    \item Assign and Constant: Represent assignments, including qubit initializations and gate parameters.
    \item Expr and BinOp: Represent mathematical expressions and quantum gate operations.
    \item Control Flow: Represent loops (\texttt{for} loops), crucial for parameterized quantum operations.
\end{itemize}

An example of a simple AST in Qiskit is as follows.
Consider the function \texttt{rx\_c} that initializes a quantum circuit, and applies a series of \texttt{rx} rotations on each of the $i \in n$ qubits with angle scaling by a factor of $\pi/i$ for $i \in [1,n]$, and returns the constructed circuit. 

\begin{lstlisting}[language=Python,caption={Qiskit example for a rx\_c module},label={rx_example}]
def rx_c(n):
    qc = QuantumCircuit(n)
    angle = pi
    for i in range(n):
        qc.rx(angle, i)
        angle /= 2
    return qc
\end{lstlisting}

\begin{figure}[h]
    \centering
    \includegraphics[width=0.7\textwidth]{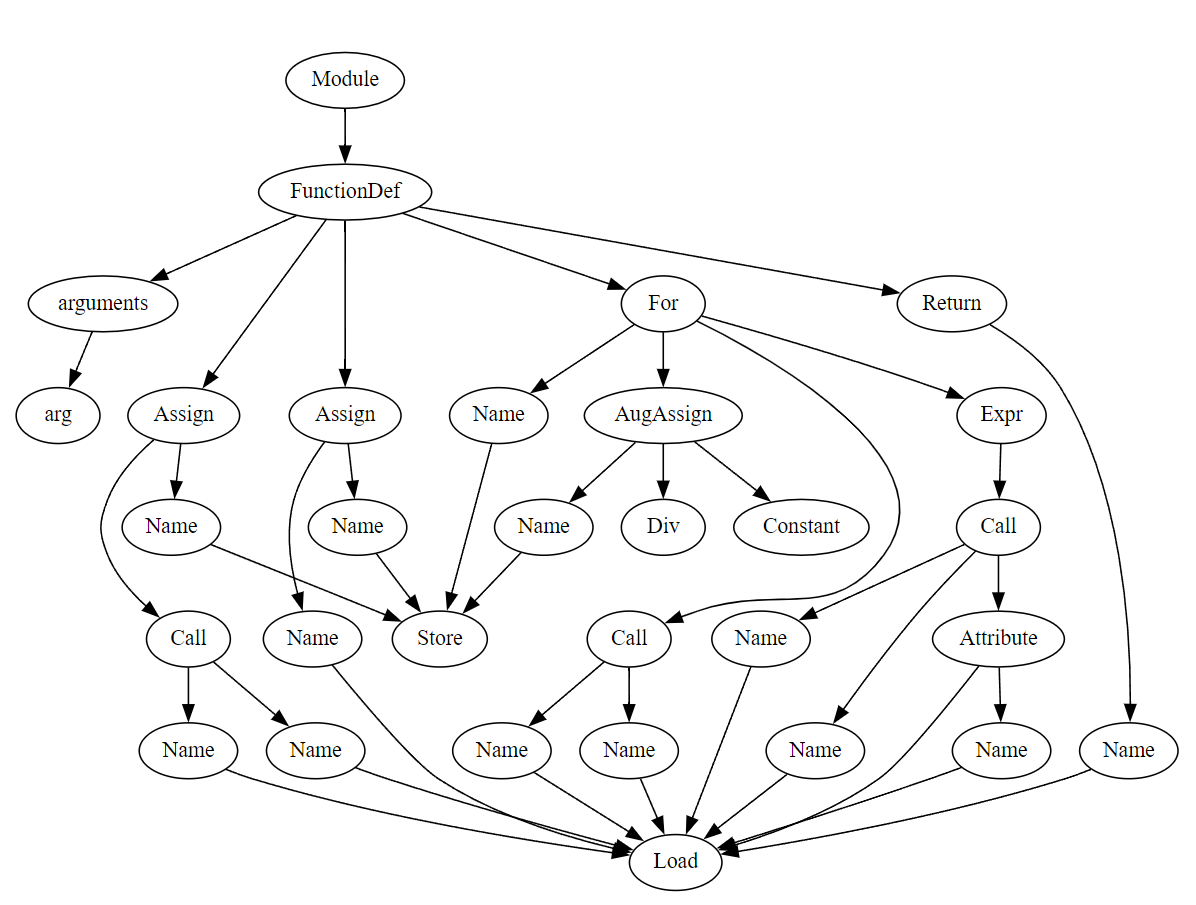}
    \caption{AST structure for a simple Qiskit function}
    \label{rx_c}
\end{figure}

The corresponding AST structure includes:
\begin{itemize}[nolistsep,noitemsep]
    \item[-] Module: The root node representing the quantum program.
    \item[-] FunctionDef: Defines the function \texttt{rx\_c} with arguments.
    \begin{itemize}[nolistsep,noitemsep]
        \item[-] Assign: Assignment and storing of the circuit object \texttt{qc = QuantumCircuit(n)}.
        \item[-] Assign: Assignment of the initial angle \texttt{angle = math.pi}.
        \item[-] For: A loop iterating over qubits.
            \begin{itemize}
                \item[-] AugAssign: Augmented assignment \texttt{angle /= 2}.
                \item[-] Call: Expression \texttt{qc.rx(angle, i)}.
            \end{itemize}
        \item[-] Load: The return statement \texttt{return qc}.
    \end{itemize}
\end{itemize}
Figure \ref{rx_c} shows the AST for the \texttt{rx\_c} function, illustrating how the AST captures the quantum circuit structure and operations.

\subsection{Initialization of the population}

By leveraging ASTs, we programmatically generate the initial population for evolution.
In each generation, a preset percentage of the individuals are also culled and regenerated.
Below are the key functions that contribute to this process.

\subsubsection{Qubit index expressions}

The \texttt{random\_expr} function generates random expressions for qubit indices using arithmetic operations, modulus, and simple variables. Parameters include:
    \begin{itemize}[nolistsep,noitemsep]
        \item depth: Number of loop variables.
        \item max\_expr\_operators: Number of binary operations in expression.
        \item var\_depth: Number of additional variables.
    \end{itemize}

For example,
\begin{itemize}[nolistsep,noitemsep]
    \item \texttt{random\_expr(0, 1, 2)}  could be a simple operation like \(n - n - n\), where \(n\) is a variable or index.
    \item \texttt{random\_expr(1, 2, 3)} might generate a more complex expression such as \(i0 - n + 3 + 2\), involving loop variables (\(i0\)), constants (3, 2), and operations.
\end{itemize}

\subsubsection{Loop structures}

Loops are essential for repeated quantum gate operations. 
The \texttt{loop\_index} function helps generate dynamic loop structures for complex quantum algorithms enabling the iterative application of quantum gates. 

Below is an example illustrating nested loops in a quantum circuit:

\begin{lstlisting}[language=Python, caption={Nested loops in quantum circuit generation}]
for i0 in range(n):
    for i1 in range(abs(i0 - n)):
        for i2 in range(abs(i0 + i1 + i1 + 1)):
            qc.crx(pi * (1 / (2 ** (i1 + n) + i1)), (n + 1) % n)
\end{lstlisting}

\begin{itemize}[nolistsep,noitemsep]
    \item First loop: The index \texttt{i0} iterates from 0 to \texttt{n}, setting the basic framework for subsequent nested loops.
    \item Second loop: Dependence on \texttt{i0} and \texttt{i1} ranges dynamically, increasing the complexity of operations performed in this layer.
    \item Third loop: The deepest layer uses both \texttt{i0} and \texttt{i1} in determining its range, illustrating an advanced level of dependency and complexity in a loop structure.
\end{itemize}

\subsubsection{Angle expressions for rotation gates}

The \texttt{random\_phase\_expr} function generates phase expressions for gates such as \texttt{rx}, \texttt{ry}, and \texttt{rz}.
This function creates an expression of the form:
\begin{equation}
    expr_{phase} = ( \pi \cdot \frac{1}{2^a + b + c}  )
\end{equation}  where \(a\) is the expression related to the number of qubits \(n\), \(b\) is the expression of the loop index $i_j$, $0<j<d$, and \(c\) are random numbers from a Gaussian distribution 
\[
    X \sim \mathcal{N}(\mu,\,\sigma^{2})\,.\mu =0  , \sigma =1
\] 

The function's only input is the depth of the current loop location, which is used to decide which symbols can be included in the expression. 
As an example, \texttt{random\_phase\_expr(2)} can generate the expression \texttt{pi * (1 / (2 ** (n + 0 + n - 0) + (i0 + 0 + 0 - n + 0)))}.

\subsubsection{Quantum gate calls}

The \texttt{generate\_gate\_call} function constructs calls to quantum gates, accommodating single, multi-qubit, and rotational gates. It uses randomly generated expressions for qubit indices and phases to define the gates' applications.
Some examples of the function \texttt{generate\_gate\_call(depth: Any, gate: Any)} are given as Table \ref{tab:gate_calls}. 
It takes the current loop depth and a specific gate type as inputs:

\begin{table}[H]
\centering
\begin{tabular}{|p{5cm}|p{11cm}|} 
\hline
Function call & Generated quantum gate operation \\
\hline
\texttt{generate\_gate\_call(2, `h')} & \texttt{qc.h((i1 - 0 - n) \% n)} \\
\hline
\texttt{generate\_gate\_call(1, `rx')} & \texttt{qc.rx(pi * (1 / (2 ** (i0 + 0 - n) + (i0 - n + n + 0))), (n - 0) \% n)} \\
\hline
\texttt{generate\_gate\_call(1, `cx')} & \texttt{qc.cx((n - 0 + n) \% n, (i0 + n - 0) \% n)} \\
\hline
\texttt{generate\_gate\_call(1, `cp')} & \texttt{qc.cp(-(pi * (1 / (2 ** (n - 0 - n) + (n - n + n + 2)))), (i0 + 0) \% n, (n - 0 - n) \% n)} \\
\hline
\end{tabular}
\caption{Examples of generating quantum gate calls using \texttt{generate\_gate\_call} function.}
\label{tab:gate_calls}
\end{table}

The \texttt{random\_qiskit\_ast\_generator} function in DeQompile assembles all these components together, constructing a complete quantum circuit. 
It takes as arguments a list of allowed operations, the maximum number of nodes, and the maximum cyclometric complexity (i.e., the levels of loop nesting).
The function iterates through AST nodes, adding gate operations and returning the final circuit. 
This approach allows for the generation of random quantum circuits with arbitrary complexity.




\subsection{Evolution operations}

The main goal is to evolve Qiskit programs (in AST representation) that can reproduce the set of QASM circuits based on the input while optimizing for resources like gate counts, node counts, etc.
The GP workflow in each generation of DeQompile after the population initialization is described in this section.



\subsubsection{Fitness evaluation} 

The fitness of each candidate quantum circuit is evaluated by comparing it to the target circuit using a set of custom evaluation metrics inspired by natural language processing (NLP). 
Similarity can broadly be classified as semantic or syntactic.
Semantic similarity can be estimated by the process fidelity between the unitary of the circuit generated via the induced program and that of the QASM target.
While semantic similarity allows generalization over syntactic differences (e.g., global phase), constructing the unitary scales exponentially with the problem size becoming restrictive beyond a few small samples.
Instead, we introduce methods that compare the syntactic structure and operational similarities between circuits. 
We use a combination of three metrics to measure the fitness of each QASM pair.

\begin{itemize}
\item Gate sequence similarity: Compares the sequences of quantum gates in the candidate and target circuits using the Levenshtein distance \cite{yujian2007normalized}. This metric calculates the minimum number of single-character edits (insertions, deletions, or substitutions) required to change one sequence to the other. The similarity score \( S_{\text{seq}} \) is defined as:
\[
S_{\text{seq}} = 1 - \sqrt{\frac{D(A, B)}{\max(|A|, |B|)}}
\]
where \( D(A, B) \) is the Levenshtein distance between the sequences \( A \) and \( B \), and \( |A| \) and \( |B| \) are their lengths. The square root emphasizes differences in larger sequences.

\item Gate frequency similarity: Assesses how similar the usage frequency of each gate type is between the candidate and target circuits. By creating frequency vectors for each circuit and calculating the cosine similarity between them, we obtain the score \( S_{\text{freq}} \):
\[
S_{\text{freq}} = \frac{\mathbf{f}_1 \cdot \mathbf{f}_2}{\|\mathbf{f}_1\| \|\mathbf{f}_2\|}
\]
where \( \mathbf{f}_1 \) and \( \mathbf{f}_2 \) are the frequency vectors of the candidate and target circuits, respectively.

\item Longest common subsequence: By finding the longest common subsequence (LCS) of lines in the QASM representations of the candidate and target circuits, we evaluate the structural similarity. The LCS score \( S_{\text{LCS}} \) is calculated as:
\[
S_{\text{LCS}} = \frac{\text{Length of LCS}}{\text{Total lines in target QASM}}
\]
A dynamic programming approach is used to efficiently compute the LCS.
\end{itemize}

The overall fitness score \( S_{\text{total}} \) for a candidate circuit is the geometric mean of the individual similarity scores.
This combined score ensures that a low similarity in any one metric significantly affects overall fitness, promoting well-rounded candidate solutions.
The total fitness of the individual includes the average $S_{\text{total}}$ for all problem sizes, and an additional quantum description complexity \cite{vitanyi2000approachesquantitativedefinitioninformation} score $S_{\text{KC}}$ as a parsimony pressure \cite{zhang1995balancing} estimated by counting the number of nodes in the AST representation of the individual:

\[
S_{\text{total}} = S_{KC} + \sum_{n} \left( S_{\text{seq}} \times S_{\text{freq}} \times S_{\text{LCS}} \right)^{1/3}
\]

These evaluation metrics focus on the structural and operational similarities between circuits, guiding the evolutionary process toward candidates that closely resemble the target circuit regarding gate sequence, usage, and overall structure with an inductive bias toward succinct representation.

\subsubsection{Genetic operations} 

Genetic programming employs genetic operations such as selection, crossover, and mutation to evolve candidate Qiskit codes over successive generations. 

Selection involves choosing individuals with better fitness scores for reproduction, allowing the propagation of superior solutions. 
DeQompile implements popular selection algorithms like roulette wheel, tournament, rank-based, random, and weighted roulette wheel.
We refer the readers to \cite{ShubingMSc} for a pedantic coverage of these methods.

Crossover combines parts of two parents' codes to explore new potential solutions.
An example is shown in Figure \ref{fig:crossover_example}.
Crossover is always done at the outmost indent level (i.e., not within a loop) to preserve the syntactic validity of the children.

The mutation operator alters parts of a single code to introduce diversity and avoid converging to suboptimal solutions.
DeQompile selects a node in the AST to mutate.
It allows either inserting a new AST by extending the body of the node or modifying the body of the node to the new AST.
Deletion is not allowed to maintain syntactic validity.

\begin{figure}[ht]
\centering

\begin{minipage}[t]{0.45\textwidth}
\begin{lstlisting}[language=Python]
# Parent 1
qc.h((n - 1) % n)
---------------- split
for i0 in range(n):
    qc.h((n - 1 - i0) % n)
    qc.h((i0 - 1) % n)
\end{lstlisting}
\end{minipage}
\hspace{0.05\textwidth} %
\begin{minipage}[t]{0.45\textwidth}
\begin{lstlisting}[language=Python]
# Parent 2
qc.h((n - 0) % n)
----------------- split
for i0 in range(n):
    qc.x((i0 + n + 1) % n)
    qc.x((i0 - n) % n)
\end{lstlisting}
\end{minipage}

\vspace{1em} %

\begin{minipage}[t]{0.45\textwidth}
\begin{lstlisting}[language=Python]
# Child 1
qc.h((n - 0) % n)
*****************  Crossover
for i0 in range(n):
    qc.x((i0 + n + 1) % n)
    qc.x((i0 - n) % n)

\end{lstlisting}
\end{minipage}
\hspace{0.05\textwidth} %
\begin{minipage}[t]{0.45\textwidth}
\begin{lstlisting}[language=Python]
# Child 2
qc.h((n - 1) % n)
******************  Crossover
for i0 in range(n):
    qc.h((n - 1 - i0) % n)
    qc.h((i0 - 1) % n)
\end{lstlisting}
\end{minipage}

\caption{An example of the effect of AST crossover on Qiskit codes}
\label{fig:crossover_example}
\end{figure}

\subsection{Improvement strategies}

To enhance the performance of the genetic decompiler, we introduced three key strategies:

\subsubsection{Random initialization of new individuals}

At each generation, a portion of the population is replaced with randomly initialized individuals to increase exploration and avoid local optima. 
The updated population for generation \(g+1\) is given by:
\[
\mathcal{P}_{g+1} = \mathcal{E}_g \cup \mathcal{N}_g
\]
where \(\mathcal{E}_g\) represents elite individuals and \(\mathcal{N}_g\) represents newly generated individuals, ensuring both diversity and retention of high-quality solutions.

\subsubsection{Annealed mutation rate}

An annealing mechanism is applied to the mutation rate, which decreases over generations to balance exploration and exploitation:
\[
m_r2(g) = \max(m_r2^0 \cdot d^g, m_r2^{\text{min}})
\]
where \(m_r2^0\) is the initial mutation rate, \(d\) is the decay factor (\(0 < d < 1\)), and \(m_r2^{\text{min}}\) is the minimum allowable rate. 
This ensures broad exploration early on and refined optimization in later stages.

These strategies improve the decompiler's ability to explore complex search spaces and converge to high-quality solutions efficiently.

\subsubsection{Simplification of expressions}

Code bloat is a known GP issue that affects the decompilation process.
To tackle this, we introduce the parsimony pressure as described in the previous section.
However, to effectively calculate this value, it is required to simplify the expressions.
This is done by using the symbolic simplification method in the SymPy package \cite{meurer2017sympy} for every generated expression (for qubits, angles, and loop limits).
Algebraic simplification \cite{moses1971algebraic} also maintains the intelligibility of the expressions in the decompiled code.

\subsection{Workflow}

The workflow of the DeQompile software is summarized in the pseudocode presented in Algorithm~\ref{alg:GP}.

\begin{algorithm}[H]
\caption{Genetic Programming Algorithm for Quantum Circuit Optimization}
\label{alg:GP}
\begin{algorithmic}[1]
\State \textbf{Initialize:} Population $\mathcal{P} \gets \text{generate\_initial\_population}(\text{pop\_size})$
\For{$\text{generation} = 1 \text{ to } \text{generations}$}
    \State Evaluate fitness $\mathcal{F}$ for each individual in $\mathcal{P}$
    \State Sort $\mathcal{P}$ by $\mathcal{F}$ in descending order
    \State Select best individual $\mathcal{I}_{\text{best}}$ with fitness $f_{\text{best}}$
    \State Initialize new population $\mathcal{P}_{\text{new}} \gets \{\mathcal{I}_{\text{best}}\}$ \Comment{Elitism}
    \For{$i = 1 \text{ to } \text{crossover\_count}$}
        \State Select parents $(\mathcal{P}_1, \mathcal{P}_2)$ using tournament selection
        \State Generate offspring $(\mathcal{C}_1, \mathcal{C}_2)$ via crossover
        \State Add $\mathcal{C}_1, \mathcal{C}_2$ to $\mathcal{P}_{\text{new}}$
    \EndFor
    \For{$i = 1 \text{ to } \text{mutation\_count}$}
        \State Mutate a random individual $\mathcal{I} \in \mathcal{P}_{\text{new}}$
    \EndFor
    \State Introduce new random individuals to maintain diversity
    \State Update $\mathcal{P} \gets \mathcal{P}_{\text{new}}$
\EndFor
\State \Return best individual $\mathcal{I}_{\text{best}}$
\end{algorithmic}
\end{algorithm}

The hyperparameter of DeQompile is listed in Table \ref{tab:hyperparameters}

\begin{table}[H]
\centering
\caption{Hyperparameters for DeQompile} \label{tab:hyperparameters}
\begin{tabular}{|l|c|p{9.5cm}|}
\hline
Hyperparameter & Default value & Description \\ \hline

algorithm\_name & N/A & The name of the quantum algorithm to be decompiled. \\ \hline
qubit\_limit & 20 & The maximum number of qubits in the generated quantum circuits. \\ \hline
generations & 100 & The number of generations the genetic algorithm will run. \\ \hline
pop\_size & 50 & The size of the population in each generation. \\ \hline
max\_length & 10 & The maximum number of operations in the generated quantum circuits. \\ \hline
crossover\_rate & 0.3 & The rate at which crossover operations occur. \\ \hline
new\_gen\_rate & 0.2 & The rate at which new random individuals are introduced to the population. \\ \hline
mutation\_rate & 0.1 & The rate at which mutation operations occur. \\ \hline
compare\_method & `l\_by\_l' & The method used to compare the generated QASM files with the target QASM files. \\ \hline
max\_loop\_depth & 2 & The maximum depth of nested loops in the generated qiskit codes. \\ \hline
selection\_method & `tournament' & The method used to select parents for crossover. \\ \hline
operations & [`h', `x', `cx'] & The list of quantum gate operations that can be used in the circuits. \\ \hline

\end{tabular}

\end{table}

\section{Challenges}

The following challenges were critical in the implementation of the DeQompile software.

\subsection{Syntactic correctness}

Syntactic correctness is a critical aspect of the quantum circuit generation process, particularly in the context of quantum genetic programming. 
Ensuring the correct syntax of quantum gates and operations is essential for both the execution of the generated circuits for fitness calculation and the final decompilation result.

\subsubsection{Decompile-time handling of runtime constraints}

At decompile time, the focus is on generating quantum circuits that adhere to the syntax of quantum programming languages, such as QASM. 
This includes ensuring that the qubit indices, gate types, and angles are well-formed according to the quantum circuit model. 
However, runtime execution introduces additional constraints, including hardware-specific requirements such as qubit connectivity and gate fidelity. 
While syntactic correctness ensures that the generated quantum circuit can be parsed and compiled into executable code, runtime correctness requires that the quantum circuit also adheres to the physical limitations of the underlying quantum processor, which are not satisfied by the decompilation.

\subsubsection{Qubit argument for multi-qubit gates}

A common issue encountered during the generation of quantum circuits is ensuring that the qubit indices fall within valid bounds of $n$. 
. To avoid syntax errors, such as an index exceeding the number of qubits, we apply a modulo operation to the expression.
\begin{equation}
 expr_{qubit} = expr_{qubit} \bmod n
\end{equation}
This approach works well for local operations (single-qubit gates). 

However, when dealing with non-local operators like the CNOT gate for two qubits or the Toffoli gate for three qubits, two/three arbitrary expressions need to be evaluated to distinct values. 
If the generated indices for both the control and target qubits are the same (i.e., CX(i, i)), the operation becomes invalid. 
This problem often arises when the randomly generated expressions for qubit indices do not properly account for the total number of qubits in the circuit. 
Since the evaluation of an expression is unknown at decompile time, such syntactic errors could not be prevented.
During fitness calculation, if such errors are encountered, the fitness for the comparison is set to $0$.


\subsubsection{Divide-by-zero error for rotation angle expressions}

A common challenge in maintaining syntactic correctness during quantum gate parameter calculations is the risk of divide-by-zero errors, particularly in gates like 
$R_y$ where parameters involve division. When the denominator approaches zero, this can lead to undefined behavior and runtime failures. To address this, one approach is to label such cases as invalid expressions and redo the sampling, ensuring that only valid parameter sets are used. This helps maintain the integrity of decompiled circuits and prevents execution issues.

\subsection{Convergence}

Convergence in evolutionary algorithms is a key factor, particularly when using genetic programming techniques to approximate a target quantum circuit. 
Achieving convergence in quantum circuit generation is challenging due to the high dimensionality of the parameter space and the complexity of quantum operations.

\subsubsection{Convergence in parameter space}

The parameter space of quantum circuits is typically large, as each quantum gate may involve several parameters (e.g., rotation angles for single-qubit gates). 
The complexity of the parameter space increases with the number of qubits and gates, making it difficult for the genetic algorithm to converge on an optimal solution. 
Convergence depends on the ability of the algorithm to effectively explore this high-dimensional space, which often requires fine-tuning the crossover and mutation operations to direct the search toward more promising regions of the solution space. 
Achieving convergence in such a complex space is non-trivial and often requires careful balance between exploration (searching new areas of the space) and exploitation (refining known good solutions) via hyperparameter tuning.

\subsubsection{Distribution of gates and expression operators}

Another challenge in achieving convergence lies in the distribution of gates and operators in the generated quantum circuits. The population of circuits evolved by the genetic algorithm may be biased toward certain types of gates, especially if those gates are overrepresented in the initial population or in the fitness evaluation function. 
For instance, an overabundance of single-qubit gates like the $U3$ gate may lead to premature convergence, where the algorithm favors these gates without exploring more diverse or complex solutions. 
To mitigate this bias and encourage the exploration of a wider variety of circuits, the distribution of gates and operators must be managed to ensure a more balanced search across the solution space. 
This can be achieved by modifying the genetic algorithm to include mechanisms that prevent over-representation of specific gates or by introducing diversity-promoting techniques such as niching or speciation.


\section{Experiments with DeQompile}

To evaluate the performance of the genetic decompiler implementation, DeQompile, we conducted a series of experiments on quantum circuit datasets, ranging from simple patterns to more complex quantum algorithms. 
This section presents the results for each dataset, along with fitness score trends and key observations.

\subsection{1-qubit ansatz}

The first set of experiments focuses on one-qubit patterns to validate the decompiler's ability to reconstruct and optimize simple circuits. 
These include single loops, multiple operations in loops, and nested loops.
These examples are motivated by common ansatz in variational algorithms like quantum alternating operator ansatz (QAOA) \cite{hadfield2019quantum}, the generalization of the quantum approximate optimization algorithm (of the same acronym, QAOA).
Many common algorithms like quantum (inverse) Fourier transformation also include similar patterns.

\begin{figure}[H]
    \centering
    \includegraphics[width=0.9\textwidth]{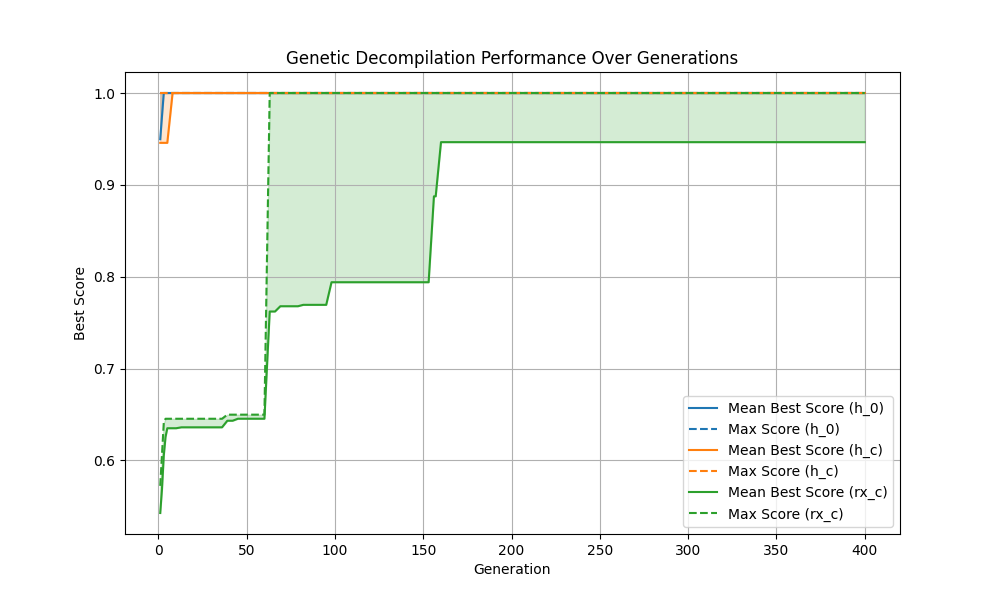}
    \caption{DeQompile performance for 1-qubit ansatz. The mean best score and maximum score across generations demonstrate steady improvement. The parameters used for generating the plot are: 
    \texttt{mutation\_rate=0.3}, 
    \texttt{pop\_size=40}, 
    \texttt{generations=100}, 
    \texttt{rep=3}, 
    \texttt{total\_qubit=20}, 
    \texttt{max\_length=10}, 
    \texttt{perform\_crossover=True}, 
    \texttt{crossover\_rate=0.3}, 
    \texttt{new\_gen\_rate=0.2}, 
    \texttt{max\_loop\_depth=2}, 
    \texttt{mutation\_rate\_2=0.5}}
    \label{fig:simple_test}
\end{figure}

\paragraph{Single loop:}
We tested circuits applying a single gate iteratively to a qubit. For example:
\begin{itemize}[nolistsep,noitemsep]
    \item Applying multiple Hadamard gate (\texttt{h\_0(n)}) to first qubit.
    \[
    \text{qc} = \prod_{i=0}^{n-1} H_0
    \]
    \item Applying the Hadamard gate (\texttt{h\_c(n)}) sequentially to all qubits.
    \[
    \text{qc} = \prod_{i=0}^{n-1} H_i
    \]
    \item Rotational gate \(R_x\) applied to each qubit with exponentially decreasing angles (\texttt{rx\_c(n)}).
    \[
    \text{qc} = \prod_{i=0}^{n-1} R_x\left(\frac{\pi}{2^i}\right)_i
    \]
\end{itemize}
The decompiler achieved a perfect fitness score of 1.0 within 30 generations, demonstrating its effectiveness in reconstructing these simple patterns.
The results of combined score over 3 trials are shown in Figure \ref{fig:simple_test}.

As the figure shows, all three simple datasets get perfectly decompiled by our decompiler, the Highest scores of duplicate experiments for each quantum circuit all achieve a combined score 1, which means all individual metrics also reach score 1, we can also confirm it by checking the final qiskit code generated by our decompiler, which matches the certain pattern of these quantum circuits.

\begin{figure}[hbt]
    \centering
    \begin{subfigure}[t]{0.5\textwidth}
        \centering
        \includegraphics[height=0.75in]{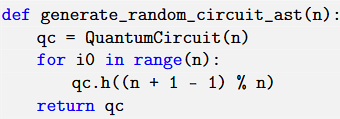}
        \caption{Best code for h\_c}
    \end{subfigure}%
    ~ 
    \begin{subfigure}[t]{0.5\textwidth}
        \centering
        \includegraphics[height=0.75in]{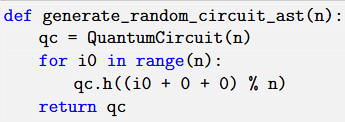}
        \caption{Best code for h\_0}
    \end{subfigure}
    ~
    \begin{subfigure}[t]{1.0\textwidth}
        \centering
        \includegraphics[height=0.9in]{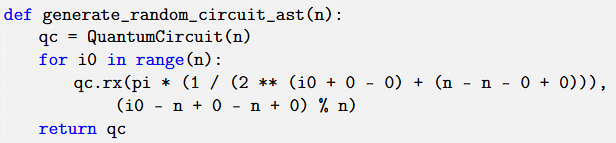}
        \caption{Best code for rx\_c}
    \end{subfigure}
    \caption{Best decompiled codes for 1-qubit ansatz}
\end{figure}

\paragraph{Multiple operations in a loop:}
We extended the experiments to circuits with multiple operations within a loop, such as alternating \(H\) and \(X\) gates. The fitness scores improved steadily, with convergence requiring slightly more generations due to increased complexity.

\paragraph{Nested loops:}
Nested loops were also evaluated, involving interdependent operations such as combining \(R_x\) gates on one qubit and \(H\) gates on another. Fitness scores displayed evolutionary jumps around 50 generations, indicating the decompiler's exploration and refinement capabilities.



\subsection{Explainability-efficiency tradeoff in gate set transpilation}

Native gates are the fundamental operations that a quantum processor can perform directly without needing further decomposition. 
Based on IBM's quantum systems \cite{ibm_native_gates}, we consider the native gates \( R_z \), \( X \), and \( \sqrt{X} \) (the square root of X), along with \( CNOT \) as a common two-qubit gate. 
These gates form the basis of quantum circuit design on these platforms and directly influence the implementation and efficiency \cite{sarkar2024yaqq, bach2023visualizing} of quantum algorithms.

In contrast to \( R_z \) and \( X \), the \( R_y \) gate (rotation around the y-axis) is not commonly included as a native gate. 
To utilize \( R_y \) operations, they must be constructed through the synthesis of available native gates, predominantly \( R_z \) and \( X \).
We tested both undecomposed and decomposed versions of \(R_y\) circuits to evaluate the decompiler's handling of circuit structure and decomposition. 
Specifically, we compared the following two scenarios as a proof of concept.

\paragraph{Undecomposed \(R_y\):} \(R_y(\theta)\) gates applied directly to all qubits.
\paragraph{Decomposed \(R_y\):} Circuits where \(R_y\) gates are expressed using \(R_x, R_z\), and \(H\) gates.
The \(R_y\) gate can be decomposed using \(RX\) and \(RZ\) gates, which are also native gates for some quantum hardware.
\[
R_y(\theta) = R_z\left(\frac{\pi}{2}\right) \cdot RX(\theta) \cdot R_z\left(-\frac{\pi}{2}\right)
\]
As an alternate common decomposition, we consider using \(H\) and \(RX\) gates
\[
R_y(\theta) = H \cdot RX(\theta) \cdot H
\]

\begin{figure}[H]
    \centering
    \includegraphics[width=0.9\textwidth]{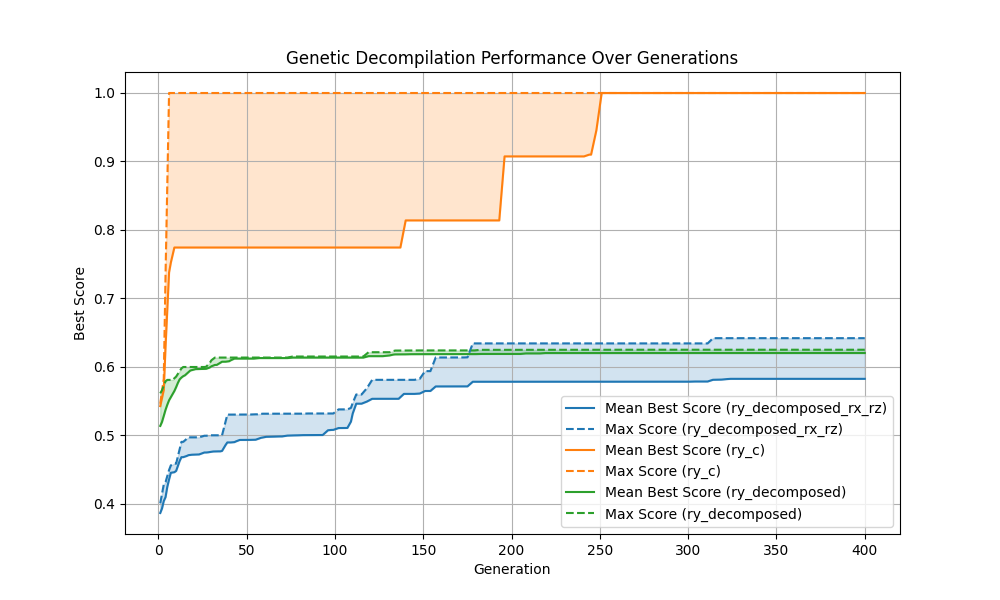}
    \caption{Fitness evolution for RY circuits for various native gate availability. Undecomposed circuits converge faster and achieve higher scores compared to decomposed versions. Here $ry\_decomposed$ means the decomposition of $r\_x$ and $h$ for the $ry\_c$ circuit, and $ry\_rx\_rz$ means the decomposition of $r\_y$ and $r\_z$ for the $r\_c$ circuit.The hyperparameter settings for the \(R_y\) gate experiments are as follows: 
\texttt{mutation\_rate=0.3}, 
\texttt{new\_gen\_rate=0.3}, \texttt{crossover\_rate=0.2}, \texttt{mutation\_rate\_2=0.99}, 
\texttt{max\_length=4}, \texttt{max\_loop\_depth=3}, \texttt{qubit\_limit=10}, \texttt{pop\_size=50}, \texttt{generations=400}, \texttt{rep=3}.}
    \label{fig:ry_decompositions}
\end{figure}

The results in Figure~\ref{fig:ry_decompositions} show that:
\begin{itemize}[nolistsep,noitemsep]
    \item Undecomposed circuits achieved perfect fitness scores within 40 generations.
    \item Decomposed circuits exhibited lower scores, with \(H\)-\(R_x\) decompositions achieving better results than \(R_x\)-\(R_z\) decompositions, which requires multiple-parameter convergence.
\end{itemize}

This experiment indicates that while decomposition improves hardware efficiency, it reduces readability, making the circuits harder to decompile.
Further aspects of explainability-efficiency tradeoff can be explored via DeQompile, especially in the context of variational algorithms \cite{morales2018variational}.

\subsection{Conventional quantum algorithms}

After conducting our decompiling experiments on simple quantum circuits, we extend the approach to more complex and commonly used quantum algorithms. 
Specifically, we selected GHZ state preparation circuits, quantum Fourier transform (QFT), and quantum phase estimation (QPE) as these exhibit a gradual increase in hierarchical and compositional complexity.
In this section, we provide a brief review of the circuit structure before presenting the decompilation results.

\paragraph{GHZ state preparation:}

\begin{figure}[ht]
\centering
\begin{adjustbox}{height=0.13\textwidth}
\begin{quantikz}
\lstick{$\ket{0}_0$} & \gate{H} & \ctrl{1} & & \ \ldots\ &  & \\
\lstick{$\ket{0}_1$} & & \gate{X} & \ctrl{1} & \ \ldots\ &  & \\
\lstick{$\ket{0}_2$} & & & \gate{X} & \ \ldots\ &  & \\
\lstick{\vdots\ } \wave&&&&&&\\
\lstick{$\ket{0}_{n-2}$} & \ghost{X} & & & \ \ldots\ & \ctrl{1}  & \\
\lstick{$\ket{0}_{n-1}$} & & & & \ \ldots\ & \gate{X} &
\end{quantikz}
\end{adjustbox}
\caption{Quantum circuit for GHZ state preparation}
\label{fig:ghz}
\end{figure}
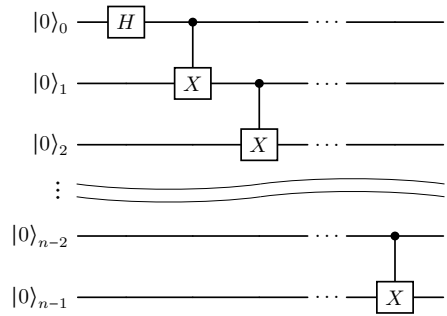

Greenberger-Horne-Zeilinger (GHZ) \cite{greenberger1989going} state is an entangled quantum state involving multiple qubits. 
It is used in quantum communication, quantum error correction, and tests of quantum mechanics. 
The circuit is shown in Figure \ref{fig:ghz}.

The key steps for constructing a GHZ state preparation circuit are:
\begin{enumerate}[nolistsep,noitemsep]
    \item Hadamard gate: Apply a Hadamard gate to the first qubit to create a superposition state.
    \item CNOT gates: Apply CNOT gates between consecutive qubits (or, alternatively, from the first qubit to other qubits) to entangle them, creating the GHZ state.
\end{enumerate}

\paragraph{Quantum Fourier transform:}

The quantum Fourier transform is a linear transformation on quantum bits, analogous to the discrete Fourier transform in classical computation \cite{lin2014shor}. 
It is a crucial component in many quantum algorithms, including Shor's algorithm for factoring. The key steps for QFT are:
\begin{enumerate}[nolistsep,noitemsep]
    \item Superposition: Apply Hadamard gates to put the qubits into a state of superposition, encoding the input in quantum parallelism.
    \item Phase rotation: Apply a series of controlled phase rotation gates to entangle the qubits and encode the Fourier transform coefficients. Each subsequent Z-rotation involves a smaller angle by a factor of 2, i.e., $\frac{\pi}{2}$, $\frac{\pi}{4}$, etc.
\end{enumerate}


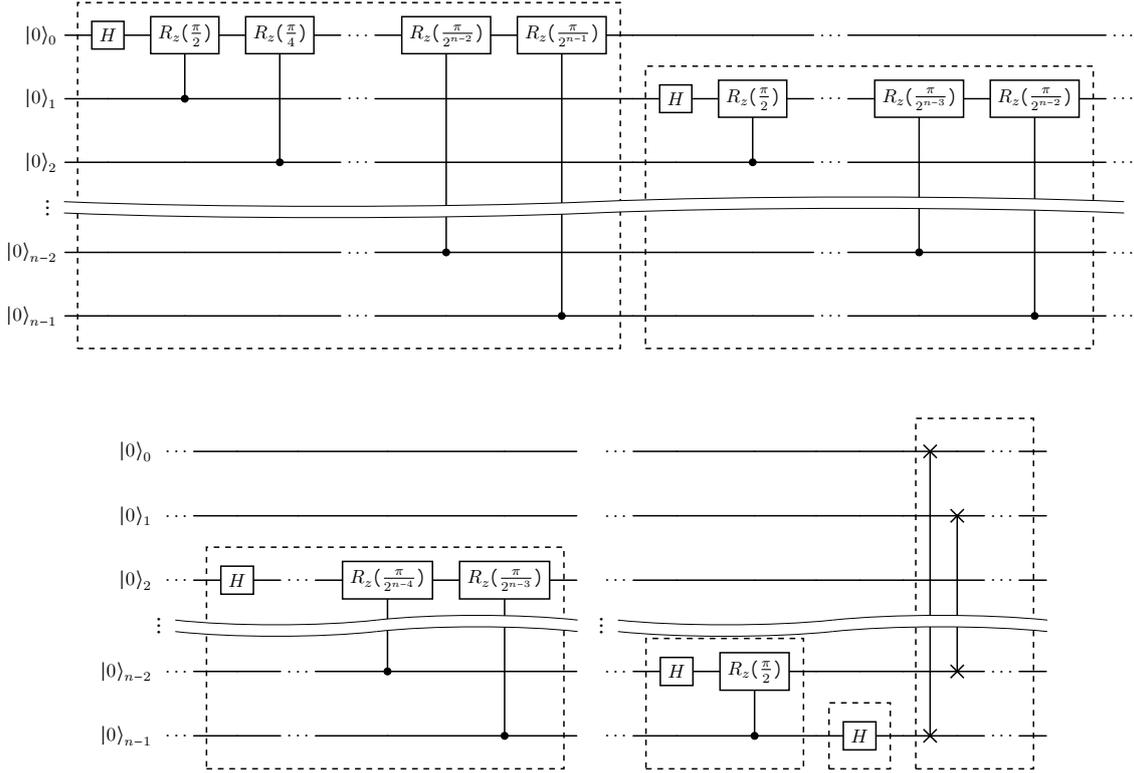
\begin{figure}[ht]
\centering
\begin{adjustbox}{height=0.15\textwidth}
\begin{quantikz}
\lstick{$\ket{0}_0$} & \gate{H}\gategroup[6,steps=6,style={dashed}]{} & \gate{R_z(\frac{\pi}{2})} & \gate{R_z(\frac{\pi}{4})} & \ \ldots\ & \gate{R_z(\frac{\pi}{2^{n-2}})} & \gate{R_z(\frac{\pi}{2^{n-1}})} & & \ghost{R_z(\frac{\pi}{2^{n-4}})} & & \ \ldots\ & & & \ \ldots\ \\
\lstick{$\ket{0}_1$} & & \ctrl{-1} &  & \ \ldots\ &  &  & & \gate{H}\gategroup[5,steps=5,style={dashed}]{} & \gate{R_z(\frac{\pi}{2})} & \ \ldots\ & \gate{R_z(\frac{\pi}{2^{n-3}})} & \gate{R_z(\frac{\pi}{2^{n-2}})} & \ \ldots\ \\
\lstick{$\ket{0}_2$} & \ghost{R_z(\frac{\pi}{2^{n-4}})} & & \ctrl{-2} & \ \ldots\ &  &  & & & \ctrl{-1} & \ \ldots\ & & & \ \ldots\ \\
\lstick{\vdots\ } \wave&&&&&&&&&&&&& \\
\lstick{$\ket{0}_{n-2}$} & \ghost{R_z(\frac{\pi}{2^{n-4}})} & & & \ \ldots\ & \ctrl{-4} &   & & & & \ \ldots\ & \ctrl{-3} & & \ \ldots\ \\
\lstick{$\ket{0}_{n-1}$} & \ghost{R_z(\frac{\pi}{2^{n-4}})} & & & \ \ldots\ & & \ctrl{-5}  & & & & \ \ldots\ & & \ctrl{-4} & \ \ldots\ 
\end{quantikz}
\end{adjustbox}\\
\vspace{0.7cm}
\begin{adjustbox}{height=0.135\textwidth}
\begin{quantikz}
\lstick{$\ket{0}_0$} \ \ldots\ & \ghost{R_z(\frac{\pi}{2^{n-4}})} &  & &  & \\
\lstick{$\ket{0}_1$} \ \ldots\ &  \ghost{R_z(\frac{\pi}{2^{n-4}})}&  &  & &  \\
\lstick{$\ket{0}_2$} \ \ldots\ & \gate{H}\gategroup[4,steps=4,style={dashed}]{}  & \ \ldots\ & \gate{R_z(\frac{\pi}{2^{n-4}})} & \gate{R_z(\frac{\pi}{2^{n-3}})} & \ghost{R_z(\frac{\pi}{2^{n-4}})} \\
\lstick{\vdots\ } \wave &&&&& \\
\lstick{$\ket{0}_{n-2}$} \ \ldots\ & \ghost{R_z(\frac{\pi}{2^{n-4}})} & \ \ldots\ & \ctrl{-2} & & \\
\lstick{$\ket{0}_{n-1}$} \ \ldots\ & \ghost{R_z(\frac{\pi}{2^{n-4}})} & \ \ldots\ & & \ctrl{-3} & 
\end{quantikz}
\end{adjustbox}
\begin{adjustbox}{height=0.152\textwidth}
\begin{quantikz}
\ldots\ & \ghost{R_z(\frac{\pi}{2^{n-4}})} &  & & & & \swap{5}\gategroup[6,steps=3,style={dashed}]{} & & \ \ldots\ & \\
\ldots\ & \ghost{R_z(\frac{\pi}{2^{n-4}})} &  & & & & & \swap{3} & \ \ldots\ & \\
\ldots\ & \ghost{R_z(\frac{\pi}{2^{n-4}})} &  & & & & & & \ \ldots\ & \\
\lstick{\vdots\ } \wave &&&&&&&&& \\
\ldots\ & \gate{H}\gategroup[2,steps=2,style={dashed}]{} & \gate{R_z(\frac{\pi}{2})} & &  & \ghost{R_z(\frac{\pi}{2^{n-4}})} & & \targX{} & \ \ldots\ & \\
\ldots\ & \ghost{R_z(\frac{\pi}{2^{n-4}})} & \ctrl{-1} & & \gate{H}\gategroup[1,steps=1,style={dashed}]{} & & \targX{} & & \ \ldots\ &
\end{quantikz}
\end{adjustbox}
\caption{Quantum circuit for quantum Fourier transform (little endian)}
\label{fig:qft}
\end{figure}

\paragraph{Quantum phase estimation:}

Quantum phase estimation \cite{kitaev1995quantum} is a fundamental algorithm used to estimate the phase (eigenvalue) introduced by a unitary operator. 
It has applications in various fields, including factoring, cryptography, and quantum chemistry. 
The key steps for QPE are:
\begin{enumerate}[nolistsep,noitemsep]
    \item State preparation: Initialize two registers: the first with qubits in superposition to act as controls, and the second with an eigenstate of the unitary operator.
    \item Controlled unitary operations: Apply controlled unitary operations that evolve the second register based on the state of the first register.
    \item Inverse quantum Fourier transform (QFT$^\dagger$): Apply the inverse QFT on the first register to convert the quantum phase information into a readable binary format.
\end{enumerate}


\begin{figure}[ht]
\centering
\begin{adjustbox}{height=0.16\textwidth}
\begin{quantikz}
\lstick{$\ket{0}_0$} & \gate{H} & \ctrl{6} & & & \ \ldots\ &  & & \gate[6]{QFT^\dagger} & \\
\lstick{$\ket{0}_1$} & \gate{H} &  & \ctrl{5} & & \ \ldots\ &  & & & \\
\lstick{$\ket{0}_2$} & \gate{H} & & & \ctrl{4} & \ \ldots\ &  & & & \\
\lstick{\vdots\ } \wave&&&&&&&&&\\
\lstick{$\ket{0}_{n-2}$} & \gate{H} & & & & \ \ldots\ & \ctrl{2}  & & &\\
\lstick{$\ket{0}_{n-1}$}  & \gate{H} & & & & \ \ldots\ &  & \ctrl{1} & & \\
\lstick{$\ket{\psi}^{\otimes m}$} & \qwbundle{m} & \gate{U^{2^0}} & \gate{U^{2^1}} & \gate{U^{2^2}} & \ \ldots\ & \gate{U^{2^{n-2}}} & \gate{U^{2^{n-1}}} & \qwbundle{m} & \ground{}
\end{quantikz}
\end{adjustbox}
\caption{Quantum circuit for quantum phase estimation. $QFT^\dagger$ refers to the inverted circuit of Figure~\ref{fig:qft}.}
\label{fig:qpe}
\end{figure}
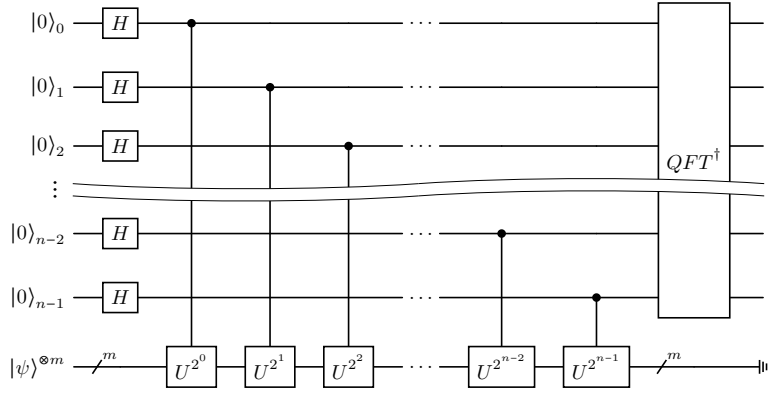

\begin{figure}[H]
    \centering
    \includegraphics[width =0.9\textwidth]{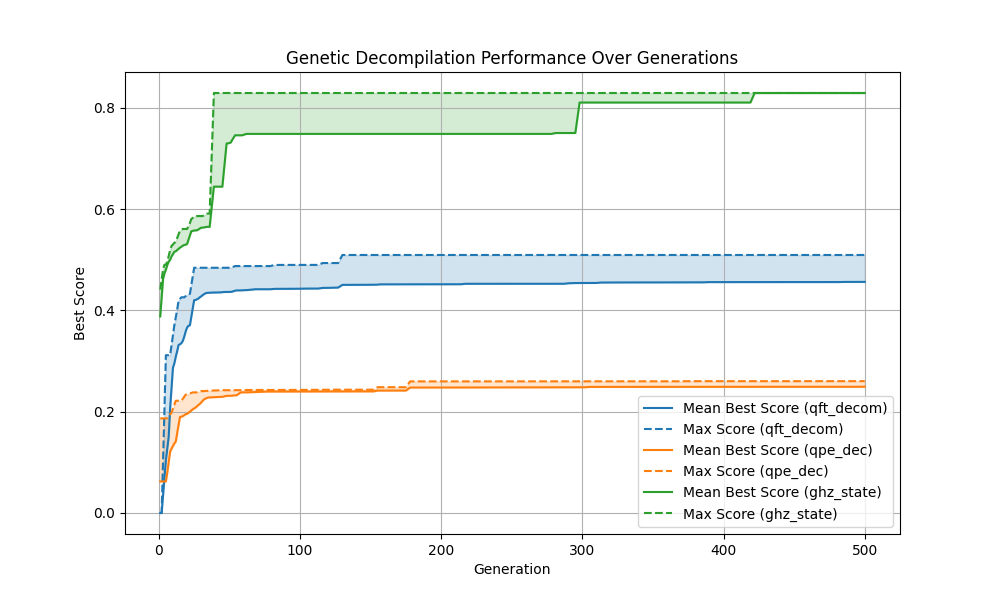}
    \caption{Genetic decompilation performance over generations. The plot shows the mean best score and the maximum score across generations for different algorithms: QFT decomposition (\texttt{qft\_decom}), QPE decomposition (\texttt{qpe\_dec}), and GHZ state preparation (\texttt{ghz\_state}). The mean best score represents the average score of the best individual across all experiments for each generation, while the max score indicates the highest score achieved among all repetitions for each generation. The hyperparameter settings are as follows: 
    \texttt{mutation\_rate=0.3}, 
    \texttt{new\_gen\_rate=0.3},
    \texttt{crossover\_rate=0.2}, \texttt{mutation\_rate\_2=0.99}, 
    \texttt{max\_length=4}, \texttt{max\_loop\_depth=3}, \texttt{qubit\_limit=10}, 
    \texttt{pop\_size=40}, \texttt{generations=500}, \texttt{rep=3}.}
    \label{com_algo}
\end{figure}

The results of the decompiling on these algorithms is shown in Figure \ref{com_algo}. 
The Qiskit code generated by the decompiler for the best individual from last generation is available in \cite{ShubingMSc}.

The plot illustrates that the GHZ circuit achieves the best decompilation result, while the QFT is easier to decompile than the QPE.
This matches the intuition of the complexity ordering among these examples.

QFT convergence was further inspected in the range of $2-10$ qubits to assess scalability. 
Smaller QFT circuits ($2-5$ qubits) achieved fitness scores above $0.9$ within $100$ generations, while larger QFT circuits ($6-10$ qubits) converged more slowly, plateauing at scores around $0.85$ after $150$ generations.
This indicates the decompiler's effectiveness for smaller patterns, while larger circuits present additional challenges due to the exponential growth in gate sequences.

\section{Conclusion}

In the current era of ubiquitous software automation, this research champions the need to distill higher-level abstractions that are both understandable to human experts and amenable to formal analysis.
Understanding the underlying algorithm behind a family of quantum circuits motivates us to develop DeQompile, a QASM to Qiskit decompiler. 
Genetic programming is employed to evolve a candidate solution over generations, using the abstract syntax tree representation of Qiskit's Python code.
We designed a software framework consisting of strategies to generate syntactically valid and expressive individuals, evolution operations, and hyperparameter tuning.
We defined a set of metrics like gate sequence frequency, gate sequence similarity, and line-by-line comparison to evaluate the fitness of the decompilation. 
The open-sourced DeQompile tool was demonstrated on a series of examples with increasing complexity, from common ansatz patterns to popular quantum algorithms.
We also demonstrate the explainability-efficiency tradeoff in quantum algorithms during native gate transpilation.
This research provides a novel explainability framework in quantum information processing.

We infer some promising future directions based on the experiments conducted.
The DeQompile tool can be extended to include measurements and control flow structures that are essential for protocols like error correction.
Our current approach to fitness calculation is based on comparing the textual QASM syntax, which fails to capture semantic similarities (e.g., commuting gates).
However, assessing semantic similarity while avoiding the manipulation of exponential-sized unitary matrices remains an open question.
It was evident that genetic programming, while being successful for symbolic regression fails to decompile or synthesize complex structures.
While further investigation into hyper-parameter tuning is imperative, it is also worthwhile to explore alternate program synthesis methods based on neural networks.
It is important to realize that while the neural network itself is not necessarily explainable (equivalent to the random mutation in our case), the overall architecture adheres to the neuro-symbolic paradigm \cite{sun2022neurosymbolic} with its distinct opportunities.
Advancements in foundational models \cite{hu2024degpt, blazek2024automated, blazek2021explainable} can augment quantum decompilation capabilities in the future.
Additionally, prior knowledge can be used to guide the convergence of the decompilation.
This can be incorporated via hierarchical reinforcement learning of a concept library \cite{trenkwalder2022automated,sarra2023discovering,kundu2024easy}.
Such an augmented framework would be equipped to decompile a suite of conventional or generated quantum circuits \cite{quetschlich2023mqtbench, apak2024ketgpt, daimon2024quantum,ruiz2024quantum}.
Finally, as an alternate use case, the DeQompile can also be used to compress quantum circuits \cite{QART, sarkar2022qksa, sarkar2022applications} and estimate the algorithmic information content \cite{steinberg2024lightcone, sarkar2022qksa, sarkar2022applications} with multiple theoretical and practical implications.



\section*{Software availability} \label{software}

The open-sourced code for the project, configuration files, output data, and plotting codes for the experiments presented in this article are available at:
\href{https://github.com/Advanced-Research-Centre/DeQompile/}{https://github.com/Advanced-Research-Centre/DeQompile/}.

\section*{Acknowledgements}

A didactic introduction and additional details of the implementation presented in this article can be found in the corresponding master thesis~\cite{ShubingMSc}.
We thank Vedran Dunjko for insightful discussions during the project planning and evaluation.
AS acknowledges funding from the Dutch Research Council (NWO) through the project ``QuTech Part III Application-based research" (project no. 601.QT.001 Part III-C—NISQ).

\section*{Author contributions}


Conceptualization, A.S.; 
Methodology, A.S., S.F. and S.X.; 
Software, S.X. and A.S.; 
Validation, S.X. and A.S.;  
Writing – Original Draft Preparation, S.X. and A.S.;  
Writing – Review \& Editing, A.S. and S.F.;  
Visualization, S.X., S.F. and A.S.; 
Supervision, A.S. and S.F.;  
Project Administration, S.F. and A.S.; 
  
\bibliographystyle{unsrt}
\bibliography{ref.bib}


\end{document}